\documentclass[12pt]{article}
\usepackage{amsmath,amssymb,epsfig,arydshln}
\usepackage{color}
\input{colordvi.tex}
\def\unit{{\relax{\rm 1\kern-.26em I}}}

\def\6#1{{\underline{#1}}}
\def\m6#1{{\underline{#1}\,}}

\newdimen\Tdim
\def\ispan{{\setbox0=\hbox{i}%
\Tdim\ht0\advance\Tdim\dp0\rule[-\dp0]{0pt}{\Tdim}}}
\def\jspan{{\setbox0=\hbox{j}%
\Tdim\ht0\advance\Tdim\dp0\rule[-\dp0]{0pt}{\Tdim}}}
\def\Tspan#1{{\setbox0=\hbox{#1}%
\Tdim\ht0\advance\Tdim\dp0\advance\Tdim.55ex\rule[-\dp0]{0pt}{\Tdim}\box0}}

\def\be{\begin{eqnarray}}
\def\ben{\begin{eqnarray*}}
\def\ee{\end{eqnarray}}
\def\een{\end{eqnarray*}}

\def\p{\partial}

\def\=:{=\hspace{-.7em}\raisebox{1.1ex}{.}\hspace{.1em}\raisebox{-0.2ex}{.} }

\newcommand {\beq}{\begin{eqnarray}}
\newcommand {\eeq}{\end{eqnarray}}
\newcommand {\non}{\nonumber\\}

\makeatletter

\renewcommand\section{\@startsection {section}{1}{\z@}%
                                   {-3.5ex \@plus -1ex \@minus -.2ex}%
                                   {2.3ex \@plus.2ex}%
                                   {\normalfont\large\bfseries}}

\renewcommand\subsection{\@startsection{subsection}{2}{\z@}%
                                     {-3.25ex\@plus -1ex \@minus -.2ex}%
                                     {1.5ex \@plus .2ex}%
                                     {\normalfont\normalsize\bfseries}}

\newcount\hour \newcount\minute
\hour=\time \divide \hour by 60
\minute=\time
\def\now{%
\ifnum \hour<13
  \ifnum \hour=0 \advance \hour by 12 \number\hour:\else \number\hour:\fi%
     \ifnum \minute<10 0\fi%
     \number\minute%
\ A.M.%
\else \advance \hour by -12 \number\hour:%
  \ifnum \minute<10 0\fi%
  \number\minute%
  \ P.M.%
\fi%
}

\makeatother

\begin{document}

% format
\baselineskip=18pt  % a la harvmac
\numberwithin{equation}{section}  % make eq labels (sec.num)
\allowdisplaybreaks  % allow page breaks in displayed eqs

% print date, time and filename
%\pagestyle{myheadings}
%\markright{{\tt \jobname.tex} -- \today{} \now}

%%%%%%%%%%%%%%%%%%%%%%%%%%%%%%%%%%%%%%%%%%%
%%%        TITLE BEGINS HERE
%%%%%%%%%%%%%%%%%%%%%%%%%%%%%%%%%%%%%%%%%%%

%% ========== title (note version) begins here ==========
%
%\vspace*{-1cm}
%\begin{center}
% {\Large\bf Title of the Document}
%\end{center}
%\vspace*{-.5cm}
%
%% ========== title (note version) ends here ==========

%% ========== title (paper version, a la harvmac) begins here ==========

\thispagestyle{empty}

% Report number
\vspace*{-2cm}
\begin{flushright}
{\tt YGHP-15-02}
%{\tt arXiv:yymm.nnnn}\\
\end{flushright}

\begin{flushright}
%MCTP-XX-XX\\
%SCIPP-XXXX/XX
\end{flushright}

\begin{center}

\vspace{-.5cm}

\vspace{0.5cm}
{\bf\Large $J$-kink domain walls and the DBI action}
\vspace*{1.5cm}

{\bf
Minoru Eto$^{a}$\footnote{\it e-mail address:
meto(at)sci.kj.yamagata-u.ac.jp} 
\vspace*{0.5cm}
}
 
$^a$ {\it {Department of Physics, Yamagata University, Kojirakawa-machi 1-4-12, 
Yamagata, Yamagata 990-8560, Japan}}

\end{center}

\vspace{1cm} \centerline{\bf Abstract} \vspace*{0.5cm}

We study $J$-kink domain walls in $D=4$ massive $\mathbb{C}P^1$ sigma model.
The domain walls are not static but stationary,  since they rotate in an internal $S^1$ space with
a frequency $\omega$ and a momentum ${\bf k}$ along the domain wall.
They are characterized by a conserved current $J_\mu = (Q,{\bf J})$, and 
are classified into magnetic ($J^2 < 0$), null ($J^2=0$), and electric ($J^2 > 0$) types.
Under a natural assumption that a low energy effective action of the domain wall is 
dual to the $D=4$ DBI action for a membrane,  
we are lead to a coincidence between the $J$-kink domain wall and 
the membrane with constant magnetic field $B$ and electric field ${\bf E}$. We also find that
$(Q, {\bf J}, \omega, {\bf k})$ is dual to $(B, {\bf E}, H, {\bf D})$ 
with $H$ and ${\bf D}$ being a magnetizing field and a displacement field, respectively.

\newpage
\setcounter{page}{1} % don't number title page

%% ========== title (paper version, a la harvmac) ends here ==========

%%%%%%%%%%%%%%%%%%%%%%%%%%%%%%%%%%%%%%%%%%%
%%%           TITLE ENDS HERE
%%%%%%%%%%%%%%%%%%%%%%%%%%%%%%%%%%%%%%%%%%%

%%%%%%%%
%%%%%%%%%%%%%%%%

%%%%%%%%%%%%%%%%

%%%%%%%%%%%%%%%%%%%%%%%%%%
\section{Introduction}

A field theory domain wall is reminiscent of a D2-brane of type IIA superstring theory.
An analogy was first pointed out for domain walls in $D=4$ supersymmetric
massive hyper-K\"ahler sigma models \cite{Abraham:1992vb}.
In the massive $T^\star \mathbb{C}P^1$ sigma model, the domain wall has collective coordinates
$Z \in \mathbb{R}$ (position in a transverse direction to the domain wall) and $\phi \in S^1$ (a Nambu-Goldstone mode
for a $U(1)$ global symmetry). Regarding the internal moduli $\phi$ as a coordinate of a ``hidden'' fifth dimension,
a low energy effective theory for the domain wall may be thought of as
an $S^1$ reduction of the $D=5$ supermembrane \cite{Abraham:1992vb}. 
Hence, the effective theory would be dual to an Abelian gauge theory,
which is quite similar to the relation between D2-brane in ten dimensions 
and M2-brane in eleven dimensions \cite{Townsend:1995af}.
Another similarity was found about the Higgs mechanism on the domain walls:  It was found that
a low energy effective field theory on $N$ domain walls top of each other
is  $U(N)$ Yang-Mills theory \cite{Arai:2012cx,Arai:2013mwa}, which is again quite similar to the D-branes.

There is a further strong evidence at qualitative level: 
In type IIA superstring theory the superstring ending on the D2-brane is a 1/4 Bogomol'nyi-Prasad-Sommerfield (BPS) state.
This can be understood as a 1/2 BPS BIon of the D=10 Dirac-Born-Inferd (DBI) 
action \cite{Callan:1997kz,Gibbons:1997xz}.
A field theory counterpart of this is a 1/4 BPS kink-lump composite solution first found in \cite{Gauntlett:2000de}
and studied later in \cite{Shifman:2002jm,Isozumi:2004vg,Sakai:2005sp,Tong:2005un,Eto:2006pg,Shifman:2007ce}.
This configuration can also be understood correctly as a $1/2$ BPS BIon of the $D=4$ DBI action \cite{Gauntlett:2000de}.
With this non-trivial coincidence between the field theory domain wall and the D2-brane in type IIA superstring theory,
it is very plausible that the low energy effective action of the domain wall is the D=4 DBI action.

The purpose of this paper is clarifying further the relation between a BPS domain wall in the massive
$\mathbb{C}P^1$ sigma model in four dimensions and a membrane in the D=4 DBI theory.
Instead of studying the relation between the kink-lump and the BIon which are three dimensionally non-trivial configuration,
we will focus on the flat domain wall and the flat membrane. A dyonic extension of the flat domain wall is so-called
the $Q$-kink domain wall \cite{Abraham:1992vb}. It is the domain wall with a conserved Noether charge $Q$. 
In Ref.~\cite{Gauntlett:2000de}, it was found that the $Q$-kink domain wall is dual
to the membrane with a constant magnetic field $B$ in the $D=4$ DBI theory. 
Now, we are lead to a simple question, what is a field theory
counterpart to the membrane with a constant electric field ${\bf E}$? Having this question in mind,
we will find new solutions, namely the $J$-kink domain walls, which
possess not only the $Q$ charge but also with a current ${\bf J}$ parallel to the domain wall.

Another perspective of this paper is finding higher derivative corrections to a low energy effective theory of
the domain wall. There are two ways: bottom-up and top-down approaches. 
The former is a conventional method requiring a brute force, see for example \cite{Eto:2012qda}: 
First, we separate fluctuations around the domain wall
background into massless and massive modes. If the massive modes are just truncated, the effective theory
includes derivatives up to quadratic order, which is the so-called the moduli approximation \cite{Manton:1981mp}. 
In order to include higher order derivative corrections, one needs to
expand the massive modes in terms of momenta and to integrate them out order by order. 
This is a straightforward task but in practice is hard to be performed.
Indeed, only a first few orders have been obtained in the literature.
On the other hand,
the latter is just assuming the effective theory of domain wall is the $D=4$ DBI action. 
As mentioned above, this is very plausible but giving a proof seems
to be difficult. Therefore, we seek non-trivial checks for this. 
One evidence is the correspondence between the kink-lump and the BIon \cite{Gauntlett:2000de}.
The results in this paper give another non-trivial evidences. Having these highly non-trivial coincidences,  now
we are quite sure that the DBI action is indeed the low energy effective action of the domain wall in the massive
$\mathbb{C}P^1$ sigma model.

The paper is organized as follows. In Sec.~\ref{sec:dw}, we construct the $J$-kink domain wall solutions
in the $D=4$ massive $\mathbb{C}P^1$ sigma model. 
Sec.~\ref{sec:DBIEM} is devoted to finding DBI counterparts to
the domain walls. In Sec.~\ref{sec:conclusion}, we conclude the results.

%%%%%%%%%%%%%%%%%%%%%%%%%%
\section{$J$-kink domain walls}
\label{sec:dw}

We are interested in topologically stable domain walls in  
the massive $\mathbb{C}P^1$ sigma model in four dimensions.
The target space $\mathbb{C}P^1$ is isomorphic to a sphere. The Lagrangian 
in terms of a standard spherical coordinate $\Theta \in [0,\pi]$ and $\Phi \in [0,2\pi)$ is given by
\beq
{\cal L} = \frac{v^2}{4}\left(-\partial_\mu\Theta\partial^\mu\Theta - \partial_\mu\Phi\partial^\mu\Phi \sin^2\Theta -
m^2 \sin^2\Theta\right).
\label{eq:lag}
\eeq
The Minkowski metric is taken to be $\eta_{\mu\nu} = (-,+,+,+)$.
Mass dimensions of the parameters $v$ and $m$ are one. We can assume $v > 0$ and $m > 0$ without loss of generality.
There are two discrete vacua at $\Theta = 0$ (the north pole) and  $\pi$ (the south pole).
Domain walls interpolating those vacua can be obtained as solutions 
for the classical equations of motion
\beq
-\p^\mu\p_\mu\Theta  + (m^2 + \p_\mu\Phi\p^\mu\Phi)\sin\Theta\cos\Theta  = 0,\quad
\p^\mu\left(\p_\mu\Phi\sin^2\Theta\right) = 0.
\label{eq:full_eom}
\eeq
The energy density is given by
\beq
{\cal E} = \frac{v^2}{4}\left[
\dot \Theta^2 + (\nabla \Theta)^2 + \left(\dot \Phi^2 + (\nabla \Phi)^2 + m^2\right) \sin^2\Theta\right].
\label{eq:ene_density}
\eeq
The Lagrangian is invariant under a $U(1)$ global transformation $\Phi \to \Phi + \alpha$.
Corresponding Noether current is given by
\beq
j_\mu = \frac{v^2}{2}\p_\mu\Phi \sin^2\Theta.
\eeq
In what follows, we will use the current density per unit area in the $x^1$-$x^2$ plane
\beq
J_\mu = (Q,{\bf J}),\quad
Q = \int dx^3\ j_0,\quad
{\bf J} = \int dx^3\ {\bf j}.
\eeq

A static domain wall  solution perpendicular to the $x^3$-axis is given by
\beq
\Phi = m \phi \ ,\quad
\Theta = 2 \arctan \left[\exp \left(\pm m (x^3-Z)\right)\right],
\label{eq:sol_static}
\eeq
with $\phi \in S^1$ and $Z \in \mathbb{R}$ are moduli parameters with mass dimension one.
We refer the solution with the upper sign as the domain wall and that with the lower sign as the anti-domain wall.
A tension of the domain wall (energy per unit area in the $x^1$-$x^2$ plane) is proportional to a topological charge $Q_{\rm T}$
\beq
T = mv^2 |Q_{\rm T}|,\qquad Q_{\rm T} \equiv \frac{1}{2}\int dx^3\ \p_3\Theta \sin\Theta = \pm 1.
\eeq
In what follows, we will take the upper sign, namely the domain wall.

A dyonic extension of the static kink domain wall was found in Ref.~\cite{Abraham:1992vb}, 
as an analogue of dyons in 3+1 dimensions \cite{Julia:1975ff}. 
It is called the $Q$-kink domain wall. 
Assuming $\Theta = \Theta(x^3)$ and $\Phi=\Phi(t)$,
the $Q$-kink domain wall solution perpendicular to the $x^3$-axis can be found 
via the Bogomol'nyi completion of the tension $\int dx^3 {\cal E}$ as
\beq
M_Q 
&=& \int dx^3\ \frac{v^2}{4}\bigg[
\left(\p_3\Theta - m\cos\alpha \sin\Theta\right)^2
+ \left(\dot\Phi + m \sin\alpha \right)^2\sin^2\Theta\non
&&\ + 2 m \p_3\Theta \cos\alpha  \sin\Theta - 2 m \dot\Phi \sin\alpha \sin^2\Theta
\bigg] \non
&\ge& \sqrt{T^2+m^2Q^2} \cos(\alpha + \delta),
\label{eq:Bogo_Q}
\eeq
where $\alpha$ is an arbitrary constant and $\tan\delta = -mQ/T$.
The inequality becomes stringent when $\alpha = - \delta$.
The bound is saturated for solutions of the BPS equations
\beq
\p_3 \Theta = m \cos\alpha \sin\Theta,\quad \dot\Phi = m\sin\alpha.
\label{eq:BPS_Q}
\eeq
The BPS tension of the $Q$-kink domain wall is given by
\beq
M_Q = \sqrt{T^2 + m^2 Q^2}.
\eeq
Writing $m\sin\alpha = \omega$ and $m\cos\alpha = \sqrt{m^2-\omega^2}$,
the solution reads
\beq
\Phi = - \omega t + m \phi,\quad
\Theta = 2 \arctan \left[\exp \left(\sqrt{m^2 -\omega^2}\, (x^3-Z)\right)\right],
\label{eq:qkink}
\eeq
where $\phi$ and $Z$ are again constants.
In terms of $\omega$, the tension and the Noether charge are expressed as
\beq
M_Q =  \frac{mT}{\sqrt{m^2-\omega^2}},\quad 
\quad Q = \frac{- v^2 \omega}{\sqrt{m^2-\omega^2}},
\label{eq:mass_Q}
\eeq
with $T=mv^2$.
Note that the $Q$-kink domain wall solution exists only for $\omega < m$.
When $\omega$ reaches at $m$, the $Q$-kink domain wall becomes infinitely broad and
the tension and the charge diverge.

For later convenience, let us rederive  the results above in another way. First, we make an ansatz  
$\Theta = \Theta(x^3)$ and $\Phi = -\omega t$. This configuration indeed solves the second 
equation of motion for $\Phi$ in Eq.~(\ref{eq:full_eom}). Thus, we are left with the unknown function $\Theta(x^3)$.
Plugging these into the Lagrangian, we get a reduced potential 
\beq
{\cal V}_{\rm red} = - {\cal L}\big|_{\Theta=\Theta(x^3),\Phi=-\omega t} 
=   \frac{v^2}{4}\left[\p_3\Theta\p^3\Theta + (m^2-\omega^2)\sin^2\Theta\right].
\label{eq:V_red}
\eeq
Let us minimize this by performing the Bogomol'nyi completion 
\beq
{\cal V}_{\rm red} &=& \frac{v^2}{4}\left[\left(\p_3\Theta - \sqrt{m^2-\omega^2} \sin\Theta\right) ^2
+ 2\sqrt{m^2-\omega^2}\p_3\Theta\sin\Theta\right] \non
&\ge& \frac{v^2}{2}\sqrt{m^2-\omega^2} \p_3\Theta \sin\Theta.
\label{eq:Bogo_red}
\eeq
The bound is saturated for
\beq
\p_3\Theta = \sqrt{m^2-\omega^2} \sin\Theta.
\eeq
This is identical to Eq.~(\ref{eq:BPS_Q}) and is solved by the solutions given in Eq.~(\ref{eq:qkink}).

Next, we generalize the $Q$-kink domain wall solution. It is easy to
verify that the following solves the equations of motion (\ref{eq:full_eom})
\beq
\Phi = - k_\mu x^\mu + m \phi,\quad
\Theta = 2 \arctan \left[\exp \left(\sqrt{m^2 + k^2}\, (x^3-Z)\right)\right],
\label{eq:new_DW}
\eeq
with $k_\mu = (\omega, {\bf k})$, ${\bf k} = (k_1,k_2,0)$, and $k^2 = k_\mu k^\mu = -\omega^2 + {\bf k}^2$. 
We will call this $J$-kink domain wall, mimicking the $Q$-kink domain wall.
The parameter should satisfy a condition 
\beq
m^2 + k^2 > 0,
\eeq
since the $J$-kink domain wall becomes infinity broad when $m^2 + k^2 = 0$.
The tension formula in terms of $k_\mu$ is given by
\beq
M_J = \frac{ v^2(m^2 + {\bf k}^2)}{\sqrt{m^2 + k^2}}.
\label{eq:mass_omega}
\eeq
When ${\bf k}=0$, the $J$-kink domain wall reduces to the $Q$-kink domain wall.
The conserved current density $J_\mu$ is related to the four momentum $k_\mu$ by
\beq
J_\mu = \frac{- v^2}{\sqrt{m^2 + k^2}} k_\mu \quad \Leftrightarrow \quad
k_\mu = \frac{-m^2}{\sqrt{T^2 - m^2 J^2}} J_\mu,
\label{eq:QJ}
\eeq
with $J^2 = J_\mu J^\mu = - Q^2 + {\bf J}^2$.
Note that $k_\mu$ and $J_\mu$ satisfies the following relation
\beq
\left(m^2 + k^2\right) \left(T^2 - m^2 J^2 \right) = m^2T^2.
\label{eq:id}
\eeq
Since $m^2 + k^2 > 0$,  $J^2$ should satisfy
\beq
m^2 J^2 < T^2
\quad\Leftrightarrow\quad
m^2{\bf J}^2 < T^2 + m^2 Q^2.
\label{eq:crit_J}
\eeq
Eliminating $\omega$ from (\ref{eq:mass_omega}), the tension formula can be expressed as
\beq
M_J = \sqrt{\left(T^2+ v^4  {\bf k}^2\right)\left(1 + \frac{Q^2}{v^4}\right)}.
\label{eq:mass_Qk}
\eeq
Further eliminating ${\bf k}$, the tension formula  in terms of  $Q$ and ${\bf J}$ is given by
\beq
M_J = \frac{T^2 + m^2Q^2}{\sqrt{T^2 - m^2 J^2}}.
\label{eq:mass_QJ}
\eeq

The $J$-kink domain walls are classified into three types according to the sign of $J^2$.
We refer to the domain walls with $J^2<0$ as magnetic type, to those with $J^2=0$ as null type, 
and to those with $J^2>0$ as electric type. A reason for the names will be explained in the next section.
The static domain wall (\ref{eq:sol_static}) is the null type
while the $Q$-kink domain wall (\ref{eq:qkink}) is the magnetic type. Since $J^2$ is a Lorentz scalar, 
the domain walls of the different types are not transformed each other by any Lorentz transformations.

Note that, however, the domain walls of the magnetic type $(J^2 < 0)$ can be obtained by boosting the $Q$-kink domain wall
($J^2 = - Q^2$). In order to see this,
let us boost the $Q$-kink domain wall  with $\tilde \omega$ given in Eq.~(\ref{eq:qkink}) with a velocity 
${\bf u} = (u_1,u_2,0)$ 
\beq
\Phi = -  \tilde \omega \frac{t + {\bf u}\cdot {\bf x}}{\sqrt{1- {\bf u}^2}} + m \phi,\quad
\Theta = 2 \arctan \left[\exp \left(\sqrt{m^2 -\tilde \omega^2}\, (x^3-Z)\right)\right].
\label{eq:Q_boost}
\eeq
Rewrite $\tilde\omega$ and ${\bf u}$ as
\beq
\frac{\tilde\omega}{\sqrt{1-{\bf u}^2}} = \omega,\quad \frac{\tilde\omega }{\sqrt{1-{\bf u}^2}}\,{\bf u} = {\bf k},
\quad 
\Rightarrow
\quad
\tilde\omega^2 = \omega^2 - {\bf k}^2 = -k^2.
\eeq
Plugging this into Eq.~(\ref{eq:Q_boost}), one reproduces the generic solutions (\ref{eq:new_DW}) with $J^2<0$.

Contrary to the magnetic type, 
the domain walls of the electric type ($J^2>0$) cannot be obtained by boosting the $Q$-kink domain wall.
Let us study the solution with $k_\mu = (0,{\bf k})$ as a representative of the electric type
\beq
\Phi = - {\bf k} \cdot {\bf x} + m \phi,\quad
\Theta = 2 \arctan \left[\exp \left(\sqrt{m^2 + {\bf k}^2}\, (x^3-Z)\right)\right].
\label{eq:J-kink}
\eeq
This has the current
\beq
J_\mu = \frac{- v^2}{\sqrt{m^2 + {\bf k}^2}} (0,{\bf k}).
\eeq
We may call this ${\bf J}$-kink domain wall.
The tension is given by
\beq
M_{\bf J} = v^2\sqrt{m^2 + {\bf k}^2} = \frac{T^2}{\sqrt{T^2 - m^2{\bf J}^2}}
\eeq
Note that since $\Phi$ is the periodic variable $\Phi \sim \Phi + 2\pi$, 
the ${\bf J}$-kink domain wall solution is periodic along $x^1$ and $x^2$ directions with period
$x^i \sim x^i + 2\pi/k^i$. Thus, this can be seen as the domain wall 
which wraps the ``compactified'' directions $x^1$ and $x^2$ with the radii $R^i = 1/k^i$.
Therefore, taking an Ansatz $\Theta=\Theta(x^3)$ and $\Phi = -{\bf k}\cdot{\bf x}$
(this solves the second equation of Eq.~(\ref{eq:full_eom})), 
the kinetic terms $(\p_i\Phi)^2$ for the $x^1$ and $x^2$ directions give the ``Kaluza-Klein" masses.
This contributes to the the reduced potential (\ref{eq:V_red}) as an additional mass term as
\beq
{\cal V}_{\rm red} = \frac{v^2}{4}\left[\p_3\Theta\p^3\Theta + (m^2 + M_{\rm KK}^2)\sin^2\Theta\right],\quad
M_{\rm KK}^2 = {\bf k}^2.
\eeq 
This is the same potential as that in Eq.~(\ref{eq:V_red}) with $m^2 - \omega^2$ being replaced by $m^2 + {\bf k}^2$.
Therefore, the Bogomol'nyi completion similar to Eq.~(\ref{eq:Bogo_red}) gives the following BPS equation
\beq
\p_3\Theta = \sqrt{m^2 + {\bf k}^2} \sin\Theta.
\eeq
As expected, this is solved by Eq.~(\ref{eq:J-kink}).
Now, the generic electric solutions can be reproduced by boosting the ${\bf J}$-kink domain wall with $\tilde{\bf k}$
given in Eq.~(\ref{eq:J-kink}) with a velocity ${\bf u} = (u_1,u_2,0)$
\beq
\Phi = - \frac{\tilde {\bf k}\cdot{\bf x} + |\tilde {\bf k}||{\bf u}|t}{\sqrt{1-{\bf u}^2}} + m \phi,\quad
\Theta = 2 \arctan \left[\exp \left(\sqrt{m^2 + \tilde {\bf k}^2}\, (x^3-Z)\right)\right].
\label{eq:J2}
\eeq
Identify $\tilde {\bf k}$ and ${\bf u}$ as
\beq
\frac{\tilde{\bf k}}{\sqrt{1-{\bf u}^2}} = {\bf k},\quad
\frac{|\tilde{\bf k}||{\bf u}|}{\sqrt{1-{\bf u}^2}} = \omega,\quad
\Rightarrow\quad
\tilde {\bf k}^2 = -\omega^2 + {\bf k}^2 = k^2.
\eeq
Plugging these into Eq.~(\ref{eq:J2}), we return to the $J$-kink domain walls of the electric type.

\begin{figure}[t]
\begin{center}
\includegraphics[width=15cm]{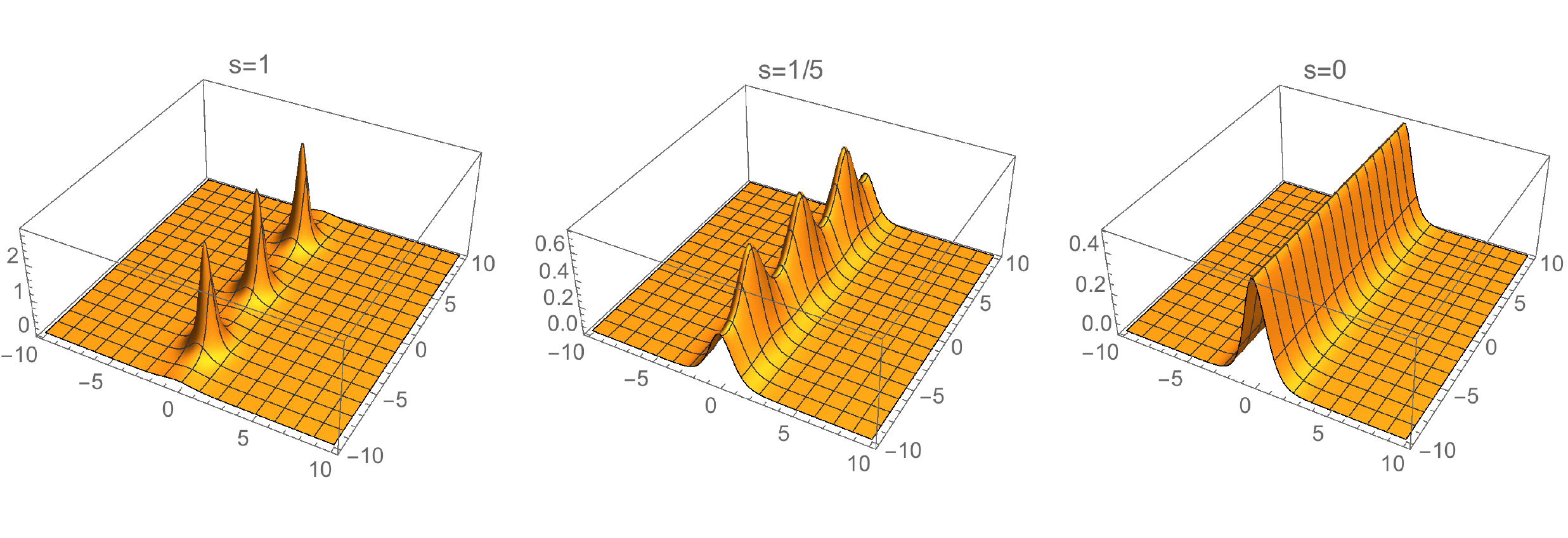}
\caption{The energy density on the $x^3$-$x^1$ plane 
for the periodic lump string configurations for $\varphi$ given in Eq.~(\ref{eq:lump_kink})
with $s=1,1/5,0$.}
\label{fig:lump_kink}
\end{center}
\end{figure}

It is interesting that the ${\bf J}$-kink domain wall seems to survive even in the massless limit $m\to0$ where the
potential term in the Lagrangian vanishes. Since the potential is absent, the non-linear sigma model becomes
the massless $\mathbb{C}P^1$ model in which whole the points on $\mathbb{C}P^1$ are vacua.
In the massless limit, instead of the domain walls, another topological soliton, the so-called lump string, appears.
In order to describe the lump strings, let us change the variable by
\beq
\varphi = e^{i\Phi}\tan\frac{\Theta}{2}.
\eeq
Then the $\mathbb{C}P^1$ Lagrangian becomes
\beq
{\cal L} = v^2 \frac{|\p_\mu\varphi|^2}{(1+|\varphi|^2)^2}.
\eeq
We consider static lump strings perpendicular to the $x^3$-$x^1$ plane.
Namely, we assume $\varphi = \varphi(x^1,x^3)$.
Introducing a complex coordinate $z = x^3+ix^1$, $\bar z = x^3 - ix^1$, $\p_z = (\p_3 - i \p_1)/2$,
and $\p_{\bar z} = (\p_3 + i \p_1)/2$,
the lump string tension can be cast into the following form
\beq
M_{\rm lump} 
&=& 2v^2 \int dx^3dx^1\ \left[
2\frac{ |\p_{\bar z}\varphi|^2}{(1+|\varphi|^2)^2} + 
\frac{|\p_z \varphi|^2 - |\p_{\bar z}\varphi|^2}{(1+|\varphi|^2)^2}
\right] \non
&\ge& 2v^2 \int dx^3 dx^1\ \frac{|\p_z \varphi|^2 - |\p_{\bar z}\varphi|^2}{(1+|\varphi|^2)^2}.
\eeq
The bound is saturated when the Bogomol'nyi equation for the lump string is satisfied
\beq
\p_{\bar z} \varphi =  0,\quad \Rightarrow \quad \varphi = \frac{P(z)}{Q(z)},
\eeq
with $P(z)$ and $Q(z)$ being polynomials in $z$ which do not have common roots. 
Then, the tension of the BPS lump string is given by
\beq
M_{\rm lump} = 2v^2\int dx^3 dx^1\ \p_z\p_{\bar z}\log (1+ |\varphi|^2) = 2\pi v^2 k,
\eeq
with  $k$ being a topological charge defined by $k \equiv \max\{\deg P,\deg Q\}$.
Now, let us consider a special solution \cite{Eto:2004rz} parametrized by two real parameters $k$ and $s$
\beq
\varphi(k,s) = (1+s)e^{-k z} - s.
\label{eq:lump_kink}
\eeq
This is periodic in $x^1$ direction with period $2\pi/k$.
Let us divide the $x^3$-$x^1$ plane into  
domains $D_n = \{(x^3,x^1)\ |\ x^3 \in (-\infty,\infty),\ x^1 \in [2n\pi/k, (2n+1)\pi/k]\}$ for $n\in \mathbb{Z}$.
Irrespective of the value of $s$, the solution (\ref{eq:lump_kink})
gives one lump string charge at each $D_n$
\beq
M_{{\rm lump}@D_n} = 2v^2 \int_{D_n}dx^3dx^1\ \p_z\p_{\bar z}\log(1+ |\varphi|^2) = 2\pi v^2.
\eeq
As is shown in Fig.~\ref{fig:lump_kink}, the lumps aligned periodically on the $x^1$ axis for $s =1$
marge into each other and
melt into the domain wall at $s=0$.
In this way, the domain wall can appear even in the massless model as a special configuration that the lump strings are aligned
periodically on a line \cite{Eto:2004rz}. Indeed, $\varphi(k,s=0)$ in Eq.~(\ref{eq:lump_kink}) is identical to the
${\bf J}$-kink domain wall with $m=0$ given in Eq.~(\ref{eq:J-kink}). Stability of the domain wall in the $m=0$ limit
is marginal, because it takes zero energy cost for transforming the domain wall into the lump strings.

Finally, we consider the null domain walls with $J^2 = 0$
\beq
\Phi = - \omega t - {\bf k}\cdot {\bf x} + m\phi,\quad 
\Theta = 2 \arctan\left[\exp\left( m (x^3 - Z)\right)\right],
\label{eq:null}
\eeq
with $\omega^2 = {\bf k}^2$. The current and the tension is given by
\beq
J_\mu = - \frac{v^2}{m}\left(\omega,{\bf k}\right),\quad
M_{\rm null} = T + \frac{m}{v^2} Q^2 = T + \frac{v^2}{m}{\bf k}^2.
\eeq
This solution can be also understood from a reduced potential as done for the $Q$-kink domain wall in Eq.~ (\ref{eq:V_red}).
Assuming $\Phi = -\omega t - {\bf k}\cdot {\bf x}$
and $\Theta = \Theta(x^3)$, the reduced potential reads
\beq
{\cal V}_{\rm red} = \p_3\Theta\p^3\Theta + m^2 \sin^2\Theta.
\eeq
This is nothing but the sine-Gordon potential and the Bogomol'nyi completion gives us
\beq
\p_3\Theta = m \sin\Theta.
\eeq
This is solved by Eq.~(\ref{eq:null}).
Let us take the domain wall of the null type with $k_\mu = (\xi, \xi,0,0)$, and boost it toward the $x^1$ direction.
It yields the following transformation
\beq
\xi \to \sqrt{\frac{1-u}{1+u}}\ \xi.
\eeq
Therefore, the static domain wall (\ref{eq:sol_static}) is obtained in the limit of $u\to 1$.

\section{The $J$-kink domain wall from the DBI action}
\label{sec:DBIEM}

In this section we will understand the domain walls with arbitrary $J_\mu$ found in the previous section from
a low energy effective action of the static domain wall ($J_\mu = 0$).

For that purpose, we start with pointing out that the $Q$-kink domain wall solution can be understood 
as a boost of the static domain wall toward the hidden ``fifth'' direction \cite{Gauntlett:2000de}.
Let us consider the massless $\mathbb{C}P^1$ sigma model in five dimensions,
\beq
S_5 = \int d^5x\  \tilde v^2
\frac{|\p_M\varphi|^2}{(1+|\varphi|^2)^2},\quad (M=0,1,2,3,4).
\eeq
The four dimensional Lagrangian (\ref{eq:lag}) can be derived through the Scherk-Schwarz (SS) dimensional reduction by
\beq
\varphi(x^\mu,w + 2\pi R_5) = e^{2\pi i m R_5}\varphi(x^\mu,w),
\label{eq:SS}
\eeq
with $w = x^4$ and $R_5$ being the radius of the fifth direction.
The mode expansion gives
\beq
\varphi(x^\mu,w) = e^{i  m w} \sum_n \varphi_n(x^\mu) e^{i\frac{n}{R_5} w}.
\eeq
In the limit of $R_5 \to 0$, all the Kaluza-Klein tower become infinitely heavy and are decoupled, 
so that we are left with the lowest mode $\varphi_0$
\beq
\varphi(x^\mu,w) = e^{i  m w}\varphi_0(x^\mu),\qquad
\varphi_0 \equiv e^{i\Phi}\tan\frac{\Theta}{2}.
\eeq
Plugging this into the fifth dimensional Lagrangian, we reproduce the massive $\mathbb{C}P^1$ sigma model as
\beq
S = \int d^4x\ \frac{v^2}{4} \left(-\p_\mu\Theta\p^\mu\Theta - \sin^2\Theta \p_\mu\Phi\p^\mu\Phi - m^2
\sin^2\Theta\right),\quad 
v^2 \equiv 2\pi R_5 \tilde v^2.
\eeq
Now, the $U(1)$ isometries of the fifth direction and the target space are linked via the SS dimensional reduction.
This implies that the moduli parameter $\phi$ appearing in $\Phi$ of the domain wall solution (\ref{eq:sol_static}) should be regarded 
as the domain wall position $w = \phi$ in the hidden fifth direction.

Let us now ``boost'' the static domain wall solution (\ref{eq:sol_static})  
toward the fifth direction. It is done by replacing the ``fifth" coordinate $W$ by
$
\phi \to \frac{\phi - ut}{\sqrt{1-u^2}}
$.
This yields a time dependence in to the domain wall solution
\beq
\Phi = m \frac{\phi - ut}{\sqrt{1-u^2}},\quad
\Theta = 2 \arctan \left[\exp \left(\pm  m (x^3-Z)\right)\right].
\eeq
Rewriting the boosted mass $m/\sqrt{1-u^2}$ as $m$ and identifying $u = \omega/m$, 
we reproduce the $Q$-kink domain wall solution (\ref{eq:qkink}).
Note that, since $u = \omega/m$ is a velocity, it is natural that the $Q$-kink domain wall solution only exist
for $\omega /m \le 1$. 
Furthermore, the tension of the $Q$-kink given in Eq.~(\ref{eq:mass_Q}) can be written as
$M_Q = T/\sqrt{1-u^2}$. This is indeed the Lorentz boosted mass formula.

In this way, it is quite natural to regard  $\phi$  in Eq.~(\ref{eq:sol_static}) to be a position of the domain wall in  the  hidden fifth direction.
Hence, a low energy effective theory of the static domain wall in the thin wall limit should be the
following Nambu-Goto type Lagrangian \cite{Gauntlett:2000de},
\beq
{\cal L}_{\rm eff} = - T \sqrt{-\det\left(\gamma_{\alpha\beta} + \p_\alpha\phi\p_\beta\phi\right)},\quad
(\alpha = 0,1,2),
\label{eq:dual_DBI}
\eeq
where $\gamma_{\alpha\beta}$ is the induced metric of the domain wall given by
\beq
\gamma_{\alpha\beta} = \eta_{\mu\nu} \frac{\p X^\mu}{\p\sigma^\alpha}\frac{\p X^\nu}{\p\sigma^\beta},
\eeq
with $\sigma^\alpha$ being a world-volume coordinate and $X^\mu$ being position of the domain wall in the
four dimensions.

Since the domain wall world-volume is $2+1$ dimensions, the effective Lagrangian (\ref{eq:dual_DBI}) 
can be dualized to the $D=4$ DBI action for a membrane by adding a BF term as
\beq
{\cal L}_{\rm eff} = 
- T \sqrt{-\det\left(\gamma_{\alpha\beta} + \p_\alpha \phi \p_\beta \phi \right)} 
+ \frac{\kappa T}{2} \epsilon_{\alpha\beta\gamma} \p^\alpha \phi F^{\beta\gamma},
\eeq
with an Abelian field strength $F_{\alpha\beta} = \p_\alpha A_\beta - \p_\beta A_\alpha$ and $\kappa$ 
is a parameter with mass dimension $-2$.
We now eliminate $\p_\alpha\phi$ from this Lagrangian by using an on-shell condition
\beq
\sqrt{-\gamma} \gamma^{\alpha\beta}\p_\beta \phi =  \kappa \sqrt{1+(\p\phi)^2} F^\alpha,\quad
F^\alpha = \frac{1}{2} \epsilon^{\alpha\beta\gamma}F_{\beta\gamma},
\label{eq:onshell}
\eeq
with $(\p\phi)^2 = \gamma^{\alpha\beta}\p_\alpha\phi\p_\beta\phi$.
Plugging this back into the Lagrangian, we find that ${\cal L}_{\rm eff}$ can be cast into the following form
\beq
{\cal L}_{\rm eff} = -T \sqrt{- \det\left(\gamma_{\alpha\beta} + \gamma^{-1} \kappa F_\alpha F_\beta\right)}.
\eeq
By using the equation $\det\left(\gamma_{\alpha\beta} + \gamma^{-1} \kappa F_\alpha F_\beta\right) = \det\left(\gamma_{\alpha\beta} + \kappa F_{\alpha\beta}\right)$,
we finally reach at the $D=4$ DBI Lagrangian for a membrane
\beq
{\cal L}_{\rm DBI} = - T \sqrt{- \det\left(\gamma_{\alpha\beta} + \kappa F_{\alpha\beta}\right)}\,.
\eeq
In the physical gauge, this is expressed as
\beq
{\cal L}_{\rm DBI} = - T \sqrt{-\det\left(\eta_{\alpha\beta} + \p_\alpha Z \p_\beta Z + \kappa F_{\alpha\beta}\right)}\,.
\label{eq:DBI}
\eeq

Since we are interested in the flat domain walls with non-zero $Q$ and ${\bf J}$, the field $Z$ is irrelevant in the following
argument. So, we will set $Z=0$ in what follows. In other words, 
we will focus on non-linear electromagnetism described by the DBI Lagrangian
\beq
{\cal L}_{\rm DBI} = - T \sqrt{1 + \frac{\kappa^2}{2}F_{\alpha\beta}F^{\alpha\beta}}.
\eeq
The equations of motion for the electromagnetic fields read
\beq
\p_\gamma \left(
\frac{\kappa^2 F^{\gamma\delta}}{\sqrt{1+ \frac{\kappa^2}{2}F_{\alpha\beta}F^{\alpha\beta}}}
\right) = 0.
\eeq
One of the simplest configurations are constant electric and magnetic fields
\beq
F_{i0} = E_i,\quad F_{12} = B.
\eeq
Note that, since the Lagrangian should be real valued, 
there is a constraint for the electric and magnetic fields
\beq
\kappa^2{\bf E}^2 \le 1 + \kappa^2 B^2.
\label{eq:crit_E}
\eeq

Let us next write down the DBI Hamiltonian. First,
the conjugate momentum (the displacement field) is given by
\beq
D_i = \frac{\p {\cal L}_{\rm DBI}}{\p E_i}
= \frac{T \kappa^2 E_i }{\sqrt{1 - \kappa^2\left({\bf E}^2 - B^2\right)}}.
\label{eq:D}
\eeq
Squaring the above equation, we get
\beq
{\bf D}^2 = \frac{T^2\kappa^4  {\bf E}^2}{1 - \kappa^2\left( {\bf E}^2 - B^2\right)}
\quad \to \quad
\kappa {\bf E} = \sqrt{\frac{1 + \kappa^2 B^2}{\kappa^2 T^2 + {\bf D}^2}}\, {\bf D}.
\label{eq:ED}
\eeq
Thus the DBI Hamiltonian is given by
\beq
{\cal H}_{\rm DBI} = {\bf D} \cdot {\bf E} - {\cal L}_{\rm DBI} = \sqrt{\left( T^2 + \frac{{\bf D}^2}{\kappa^2}\right) \left(1 + \kappa^2 B^2\right)}.
\label{eq:H_DBI}
\eeq

Now we are ready to compare the constant electric and magnetic fields on the membrane in the DBI theory 
with the $J$-kink domain wall  studied in Sec.~\ref{sec:dw}.
A natural identification is given by the on-shell condition (\ref{eq:onshell}), as follows.
Since we set $Z=0$, the induced metric in the physical gauge is $\gamma_{\alpha\beta} = \eta_{\alpha\beta}$.
Hence,
from the on-shell condition with $\phi = -(\omega t + {\bf k} \cdot {\bf x})/m$, see Eq.~(\ref{eq:new_DW}), we find
the following relation between $\{B,{\bf E}\}$ and $\{\omega,{\bf k}\}$
\beq
\kappa B = \frac{- \omega}{\sqrt{m^2 + k^2}},\quad 
\kappa E_i = \frac{-\epsilon_{ij}k^j}{\sqrt{m^2 + k^2}}.
\label{eq:BE_wk}
\eeq
Comparing this with Eq.~(\ref{eq:QJ}), we are lead to the following  identification
\beq
\kappa B  = \frac{Q}{v^2},\quad 
\kappa E_i  = -\frac{\epsilon_{ij}J^j}{v^2}.
\label{eq:identification_EBQJ}
\eeq
The first relation between the membrane with the constant magnetic field $B$ and the
$Q$-kink domain wall was found in Ref.~\cite{Gauntlett:2000de}.
Eq.~(\ref{eq:identification_EBQJ}) is a generalization of this: Namely,
the membrane with the constant magnetic and electric fields in the DBI theory corresponds
to the $J$-kink domain wall  via the identification (\ref{eq:identification_EBQJ}).

As the $J$-kink domain walls are classified into three types according to the sign of the Lorentz scalar $J^2$,
the membrane in the DBI theory can be classified into three types by the sign of a Lorentz scalar $B^2 - {\bf E}^2$.
The membranes with $B^2 - {\bf E}^2 > 0$ can be obtained by a Lorentz transformation of the membrane with $(B,{\bf E}) = (B,{\bf 0})$.
Similarly, the membranes with $B^2 - {\bf E}^2 < 0$ can be gotten from the membrane with $(B,{\bf E}) = (0,{\bf E})$.
This is a reason why we christened the $J$-kink domain wall with $J^2>0$ ($J^2<0$) the magnetic (electric) type.

The identification (\ref{eq:identification_EBQJ}) connects many quantities of the $J$-kink domain wall and
the membrane with the constant electric and magnetic fields: For example, the constraint to $Q$ and ${\bf J}$ 
given in Eq.~(\ref{eq:crit_J}) are dual to the constraint to $B$ and ${\bf E}$ given in Eq.~(\ref{eq:crit_E}).
In addition, the condition follows from Eq.~(\ref{eq:BE_wk}) 
\beq
\left(m^2 + k^2\right)\left(1 - \kappa^2(-B^2 + {\bf E}^2)\right) = m^2,
\label{eq:id_BE}
\eeq
is dual to Eq.~(\ref{eq:id}).
The constraint Eqs.~(\ref{eq:crit_J}) for $\{Q,{\bf J}\}$ is also dual to the constraint (\ref{eq:crit_E})
for $\{B,{\bf E}\}$.

Finally, let us verify the tension formulae of the $J$-kink domain wall and the membrane.
By using the identity (\ref{eq:id_BE}) and $T = mv^2$, 
the displacement field ${\bf D}$ in Eq.~(\ref{eq:D}) can be written as
\beq
\frac{D_i}{\kappa} = \frac{T}{\sqrt{1-\kappa^2\left( - B^2 + {\bf E}^2\right)}} \frac{-\epsilon_{ij}k^j}{\sqrt{m^2 + k^2}} 
=  v^2 \left(-\epsilon_{ij}k^j\right).
\eeq
Plugging this into the Hamiltonian (\ref{eq:H_DBI}), we reach at the tension formula of 
the $J$-kink domain wall (\ref{eq:mass_QJ}) expressed by $Q$ and ${\bf k}$.
Furthermore, combining Eqs.~(\ref{eq:ED}) and (\ref{eq:H_DBI}), 
the Hamiltonian of  the membrane is written in terms of ${\bf E}$ and $B$ as
\beq
{\cal H}_{\rm DBI} = \frac{T\left(1+ \kappa^2B^2\right)}{\sqrt{1-\kappa^2\left( - B^2 + {\bf E}^2\right)}}.
\label{eq:ene_DBI}
\eeq
With the identification Eq.~(\ref{eq:identification_EBQJ}), this is equal to the tension formula (\ref{eq:mass_QJ})
of the $J$-kink domain wall in terms of $Q$ and ${\bf J}$.

In order to complete the identification, let us find a counterpart to the magnetizing field $H$ (conjugate of $B$) by
\beq
\frac{H}{\kappa} = - \frac{1}{\kappa}\frac{\p {\cal L}_{\rm DBI}}{\p B}
= \frac{T \kappa B}{\sqrt{1 - \kappa^2\left( - B^2 + {\bf E}^2 \right)}} = -  v^2 \omega.
\eeq
Thus, the correspondence between the $J$-kink domain wall in the massive $\mathbb{C}P^1$ sigma model and 
the membrane with the constant electromagnetic field in the $D=4$ DBI theory is summarized as
\beq
\frac{Q}{v^2} = \kappa B,\quad \frac{J_i}{v^2} = \kappa \epsilon_{ij}E^j,\quad 
v^2 k_i = \frac{\epsilon_{ij}D^j}{\kappa},\quad v^2 \omega = - \frac{H}{\kappa}.
\eeq
The expression becomes simpler if we choose $\kappa = 1/v^2$.

\section{Concluding remarks}
\label{sec:conclusion}

In this paper, we studied the $J$-kink domain wall in the massive $\mathbb{C}P^1$ sigma model in four dimensions,
which is a generalization of the $Q$-kink domain wall \cite{Abraham:1992vb}. The $J$-kink domain walls are classified
into the three types: the magnetic type ($J^2<0$), the null type ($J^2 = 0$), and the electric type ($J^2>0$).
The domain walls of the magnetic type can be obtained by boosting the $Q$-kink domain wall while those
of the electric type can be gotten by boosting the ${\bf J}$-kink domain wall. The domain walls of the null type
includes the static domain wall ($J_\mu = 0$). The generic domain walls of the null type reach at the static domain
wall if they are boosted along the domain wall with the speed of light.

We explicitly showed that the $Q$-kink domain wall can be regarded as the domain wall which is boosted 
toward the hidden fifth direction. This fact strongly suggests that the low energy effective theory of the domain wall
in the thin wall limit is dual to the $D=4$ DBI action for the membrane \cite{Abraham:1992vb,Gauntlett:2000de}. 
Assuming it is indeed the case, we found that the membranes with the constant electric and magnetic fields are
counterpart to the $J$-kink domain walls. The dictionary is $(Q,{\bf J},\omega,{\bf k}) \Leftrightarrow (B,{\bf E},H,{\bf D})$.
With this dictionary at hand, we found that many quantities, for example, the tension formulae of the domain wall
and the membrane precisely coincide. These non-trivial coincidences together with the another coincidence between
the kink-lump and the BIon \cite{Gauntlett:2000de} tell that the low energy effective theory of the domain wall
in the massive $\mathbb{C}P^1$ sigma model is the $D=4$ DBI action.

Another perspective of achievement of this paper is specifying higher derivative corrections to the low energy 
effective action in the moduli approximation (MA). The effective theory in MA  can be obtained 
by promoting the zero modes $Z$ and $\phi$ to be fields on the world-volume of the domain wall. By a standard
procedure, the effective Lagrangian can be found as
\beq
{\cal L}_{\rm eff}^{\text{MA}} = -T - \frac{T}{2}\left(\p_\alpha Z\p^\alpha Z + \p_\alpha \phi \p^\alpha\phi\right).
\label{eq:lag_MA}
\eeq
This is nothing but the first two terms in expansion  
of the Lagrangian (\ref{eq:dual_DBI}) in the physical gauge in terms of the derivative $\p_\alpha$.
A solution is given by $Z=0$ and $\phi = - (\omega t + {\bf k}\cdot {\bf x})/m$, which correctly reproduces $\Phi$ of
the $J$-kink domain wall solution (\ref{eq:new_DW}). However, since MA is valid only
for a small $\p_\alpha$, only the solutions with small $\omega/m$ and $|{\bf k}|/m$ can be described by
Eq.~(\ref{eq:lag_MA}). In fact, the tension from Eq.~(\ref{eq:lag_MA}) is 
\beq
M_J^{\text{MA}} = T + \frac{T}{2m^2}\left(\omega^2 + {\bf k}^2\right),
\eeq
which is consistent with the generic tension formula (\ref{eq:mass_omega}) only for small $\omega$ and $|{\bf k}|$.
In order to reproduce the $J$-kink domain walls with bigger $\omega$ and $|{\bf k}|$, we should go beyond MA.
Namely, we have to taking into account higher derivative corrections.
However, as is mentioned in the Introduction, finding the higher derivative corrections to MA is not an easy task.
In this paper, in order to keep ourself away from being involved into such a complicated work, we jumped to the DBI
action which gives the correct tension formula to the all order of $\omega/m$ and $|{\bf k}|/m$.
A similar strategy was recently applied to the dyonic non-Abelian vortex, and a low energy 
effective Lagrangian including higher derivative corrections to the all order was proposed \cite{Eto:2014gya}.

There are several future directions. 
The most interesting point would be generalizing the results of this paper to multiple domain walls. 
In this work, we considered single domain wall in the $\mathbb{C}P^1$ sigma model, 
and found the correspondence to the Abelian DBI theory. 
As is well-known, $N$ BPS domain walls exist in the massive $\mathbb{C}P^{N}$ sigma model.   When the $N$ domain
walls are top of each other, a non-Abelian symmetry would emerge and a non-Abelian extension of the DBI action 
might appear as a counterpart.
Another direction is searching other $J$-solitons of known $Q$-solitons, like $Q$-lumps \cite{Abraham:1991ki} 
and dyonic non-Abelian vortices \cite{Collie:2008za} in higher dimensions.

%%%%%%%%%%%%%%
\section*{Acknowledgments}

This work is supported by Grant-in-Aid for Scientific Research No. 26800119. 
The author thanks the organizers of the workshop, string theory in future
- on the occasion of closing of the Hashimoto mathematical physics group at RIKEN,
where this work was initiated. The author thanks Koji Hashimoto and Ryuske Endo for fruitful discussion.
%\begin{appendix}
%%%%%%%%%%%%%%%%
%\section{blah}
%
%\end{appendix}
%%%%%%%%%%%%%%%%%%%%%%%%%%%%%%%%%%%%%%%%%%%%%%%%%%%%%%%%%%%%


\begin{thebibliography}{100}

%\cite{Abraham:1992vb}
\bibitem{Abraham:1992vb} 
  E.~R.~C.~Abraham and P.~K.~Townsend,
  ``Q kinks,''
  Phys.\ Lett.\ B {\bf 291}, 85 (1992).
  %%CITATION = PHLTA,B291,85;%%
  %120 citations counted in INSPIRE as of 06 mar 2015

%\cite{Townsend:1995af}
\bibitem{Townsend:1995af} 
  P.~K.~Townsend,
  ``D-branes from M-branes,''
  Phys.\ Lett.\ B {\bf 373}, 68 (1996)
  [hep-th/9512062].
  %%CITATION = HEP-TH/9512062;%%
  %445 citations counted in INSPIRE as of 06 Mar 2015

%\cite{Arai:2012cx}
\bibitem{Arai:2012cx} 
  M.~Arai, F.~Blaschke, M.~Eto and N.~Sakai,
  ``Matter Fields and Non-Abelian Gauge Fields Localized on Walls,''
  PTEP {\bf 2013}, 013B05 (2013)
  [arXiv:1208.6219 [hep-th]].
  %%CITATION = ARXIV:1208.6219;%%
  %3 citations counted in INSPIRE as of 06 Mar 2015

%\cite{Arai:2013mwa}
\bibitem{Arai:2013mwa} 
  M.~Arai, F.~Blaschke, M.~Eto and N.~Sakai,
  ``Stabilizing matter and gauge fields localized on walls,''
  PTEP {\bf 2013}, no. 9, 093B01 (2013)
  [arXiv:1303.5212 [hep-th]].
  %%CITATION = ARXIV:1303.5212;%%

%\cite{Callan:1997kz}
\bibitem{Callan:1997kz} 
  C.~G.~Callan and J.~M.~Maldacena,
  %``Brane death and dynamics from the Born-Infeld action,''
  Nucl.\ Phys.\ B {\bf 513}, 198 (1998)
  [hep-th/9708147].
  %%CITATION = HEP-TH/9708147;%%
  %479 citations counted in INSPIRE as of 06 mar 2015

%\cite{Gibbons:1997xz}
\bibitem{Gibbons:1997xz} 
  G.~W.~Gibbons,
  ``Born-Infeld particles and Dirichlet p-branes,''
  Nucl.\ Phys.\ B {\bf 514}, 603 (1998)
  [hep-th/9709027].
  %%CITATION = HEP-TH/9709027;%%
  %363 citations counted in INSPIRE as of 06 mar 2015

%\cite{Gauntlett:2000de}
\bibitem{Gauntlett:2000de} 
  J.~P.~Gauntlett, R.~Portugues, D.~Tong and P.~K.~Townsend,
  ``D-brane solitons in supersymmetric sigma models,''
  Phys.\ Rev.\ D {\bf 63}, 085002 (2001)
  [hep-th/0008221].
  %%CITATION = HEP-TH/0008221;%%
  %94 citations counted in INSPIRE as of 04 Mar 2015

%\cite{Shifman:2002jm}
\bibitem{Shifman:2002jm} 
  M.~Shifman and A.~Yung,
  ``Domain walls and flux tubes in N=2 SQCD: D-brane prototypes,''
  Phys.\ Rev.\ D {\bf 67}, 125007 (2003)
  [hep-th/0212293].
  %%CITATION = HEP-TH/0212293;%%
  %79 citations counted in INSPIRE as of 06 mar 2015

%\cite{Isozumi:2004vg}
\bibitem{Isozumi:2004vg} 
  Y.~Isozumi, M.~Nitta, K.~Ohashi and N.~Sakai,
  ``All exact solutions of a 1/4 Bogomol'nyi-Prasad-Sommerfield equation,''
  Phys.\ Rev.\ D {\bf 71}, 065018 (2005)
  [hep-th/0405129].
  %%CITATION = HEP-TH/0405129;%%
  %142 citations counted in INSPIRE as of 06 mar 2015

%\cite{Sakai:2005sp}
\bibitem{Sakai:2005sp} 
  N.~Sakai and D.~Tong,
  ``Monopoles, vortices, domain walls and D-branes: The Rules of interaction,''
  JHEP {\bf 0503}, 019 (2005)
  [hep-th/0501207].
  %%CITATION = HEP-TH/0501207;%%
  %65 citations counted in INSPIRE as of 06 Mar 2015

%\cite{Tong:2005un}
\bibitem{Tong:2005un} 
  D.~Tong,
  ``TASI lectures on solitons: Instantons, monopoles, vortices and kinks,''
  hep-th/0509216.
  %%CITATION = HEP-TH/0509216;%%
  %190 citations counted in INSPIRE as of 06 Mar 2015

%\cite{Eto:2006pg}
\bibitem{Eto:2006pg} 
  M.~Eto, Y.~Isozumi, M.~Nitta, K.~Ohashi and N.~Sakai,
  ``Solitons in the Higgs phase: The Moduli matrix approach,''
  J.\ Phys.\ A {\bf 39}, R315 (2006)
  [hep-th/0602170].
  %%CITATION = HEP-TH/0602170;%%
  %215 citations counted in INSPIRE as of 06 Mar 2015

%\cite{Shifman:2007ce}
\bibitem{Shifman:2007ce} 
  M.~Shifman and A.~Yung,
  ``Supersymmetric Solitons and How They Help Us Understand Non-Abelian Gauge Theories,''
  Rev.\ Mod.\ Phys.\  {\bf 79}, 1139 (2007)
  [hep-th/0703267].
  %%CITATION = HEP-TH/0703267;%%
  %163 citations counted in INSPIRE as of 06 Mar 2015

%\cite{Eto:2012qda}
\bibitem{Eto:2012qda} 
  M.~Eto, T.~Fujimori, M.~Nitta, K.~Ohashi and N.~Sakai,
  ``Higher Derivative Corrections to Non-Abelian Vortex Effective Theory,''
  Prog.\ Theor.\ Phys.\  {\bf 128}, 67 (2012)
  [arXiv:1204.0773 [hep-th]].
  %%CITATION = ARXIV:1204.0773;%%
  %13 citations counted in INSPIRE as of 07 mar 2015

%\cite{Manton:1981mp}
\bibitem{Manton:1981mp} 
  N.~S.~Manton,
  ``A Remark on the Scattering of BPS Monopoles,''
  Phys.\ Lett.\ B {\bf 110}, 54 (1982).
  %%CITATION = PHLTA,B110,54;%%
  %494 citations counted in INSPIRE as of 31 Mar 2015


%\cite{Julia:1975ff}
\bibitem{Julia:1975ff} 
  B.~Julia and A.~Zee,
  ``Poles with Both Magnetic and Electric Charges in Nonabelian Gauge Theory,''
  Phys.\ Rev.\ D {\bf 11}, 2227 (1975).
  %%CITATION = PHRVA,D11,2227;%%
  %478 citations counted in INSPIRE as of 26 mar 2015
  
%\bibitem{vilenkin2000cosmic}
% A.~Vilenkin, and E~P.~S~Shellard,
% ``Cosmic strings and other topological defects,"
%  Cambridge University Press.
  
%\cite{Eto:2004rz}
\bibitem{Eto:2004rz} 
  M.~Eto, Y.~Isozumi, M.~Nitta, K.~Ohashi and N.~Sakai,
  ``Instantons in the Higgs phase,''
  Phys.\ Rev.\ D {\bf 72}, 025011 (2005)
  [hep-th/0412048].
  %%CITATION = HEP-TH/0412048;%%
  %97 citations counted in INSPIRE as of 31 Mar 2015
  

%\cite{Eto:2014gya}
\bibitem{Eto:2014gya} 
  M.~Eto and Y.~Murakami,
  ``Dyonic non-Abelian vortex strings in supersymmetric and non-supersymmetric theories: tensions and higher derivative corrections,''
  arXiv:1412.7892 [hep-th].
  %%CITATION = ARXIV:1412.7892;%%

%\cite{Abraham:1991ki}
\bibitem{Abraham:1991ki} 
  E.~Abraham,
  ``Nonlinear sigma models and their Q lump solutions,''
  Phys.\ Lett.\ B {\bf 278}, 291 (1992).
  %%CITATION = PHLTA,B278,291;%%
  %35 citations counted in INSPIRE as of 31 Mar 2015

%\cite{Collie:2008za}
\bibitem{Collie:2008za} 
  B.~Collie,
  ``Dyonic Non-Abelian Vortices,''
  J.\ Phys.\ A {\bf 42}, 085404 (2009)
  [arXiv:0809.0394 [hep-th]].
  %%CITATION = ARXIV:0809.0394;%%
  %17 citations counted in INSPIRE as of 31 Mar 2015

\end{thebibliography}
\end{document}